\documentclass[aps,prl,twocolumn,showpacs,showkeys,superscriptaddress]{revtex4-1}

\usepackage{graphicx}
\usepackage{multirow}

\begin{document}



\title{Anisotropic two-dimensional electron gas at SrTiO$_3$ (110) protected by its native overlayer}

\author{Zhiming Wang}

\affiliation{Institute of Applied Physics, Vienna University of Technology, Vienna, Austria.}

\author{Zhicheng Zhong}

\affiliation{Institute of Solid State  Physics, Vienna University of Technology, Vienna, Austria.}

\author{Xianfeng Hao}

\author{Stefan Gerhold}

\author{Bernhard St\"{o}ger}

\author{Michael Schmid}

\affiliation{Institute of Applied Physics, Vienna University of Technology, Vienna, Austria.}

\author{Jaime S\'{a}nchez-Barriga}

\author{Andrei Varykhalov}

\affiliation{BESSY, Albert-Einstein-Str. 15, D-12489 Berlin, Germany.}

\author{Cesare Franchini}

\affiliation{Faculty of Physics and Center for Computational Material Science, University of Vienna, Vienna, Austria.}

\author{Karsten Held}

\affiliation{Institute of Solid State  Physics, Vienna University of Technology, Vienna, Austria.}

\author{Ulrike Diebold}

\affiliation{Institute of Applied Physics, Vienna University of Technology, Vienna, Austria.}









\maketitle

\textbf{Two dimensional electron gases (2DEGs) at oxide heterostructures are
attracting considerable attention, as these might substitute conventional
semiconductors for novel electronic devices \cite{Mannhart:sci10}.  Here we present a minimal
set-up for such a 2DEG -- the SrTiO$_3$ (110)-(4~$\times$~1) surface, natively terminated with
one monolayer of chemically-inert titania.  Oxygen vacancies induced by
synchrotron radiation migrate underneath this overlayer, this leads to a
confining potential and electron doping such that a 2DEG develops. Our
angular resolved photoemission spectroscopy (ARPES) and theoretical
results show that confinement along (110) is strikingly different from a
(001) crystal orientation. In particular the quantized subbands show a
surprising ``semi-heavy'' band, in contrast to the analogue in the bulk, 
and a high electronic anisotropy. This anisotropy and even the effective mass of the (110) 2DEG is tunable by doping, offering a high flexibility to engineer the properties of this system.
} 

With ongoing miniaturization, conventional Si-based semiconductors are
reaching their limits.  Current technology is already based on a
hybrid between Si and a HfO$_2$-based high-$k$ dielectric. 
Another transition metal oxide, SrTiO$_3$,  has a ten times
larger dielectric constant and would allow for even smaller structures.
Against this background the  2DEG observed 
in oxide heterostructures such as LaAlO$_3$/SrTiO$_3$ \cite{Ohtomo:nat04} offers a promising alternative, not only for electronics at the nanoscale \cite{Cen:sc09} but also because of the possibility of spin-polarized   \cite{Brinkman:natm07} and superconducting \cite{Thiel:sc06,Reyren:sc07} currents. 
An even simpler set-up is to create a
2DEG directly at SrTiO$_3$. Recently this was achieved by irradiating a (001) surface \cite{Santander:nat11,Meevasana:natm11} with synchrotron radiation, albeit the origin of the resulting 2DEG is still under debate  \cite{Santander:nat11,Meevasana:natm11,Plumb:arxiv13}. This system  has two major drawbacks:
(i) it is 
unstable  outside an ultra high vacuum (UHV) chamber
and (ii) the (001) surface has no unique surface termination, as TiO$_2$  and  SrO
terraces may develop, and the surface structure strongly depends on sample treatment and history \cite{Bonnell:rpp08}.

 \begin {figure}[t]
 \includegraphics [width=3.4 in,clip] {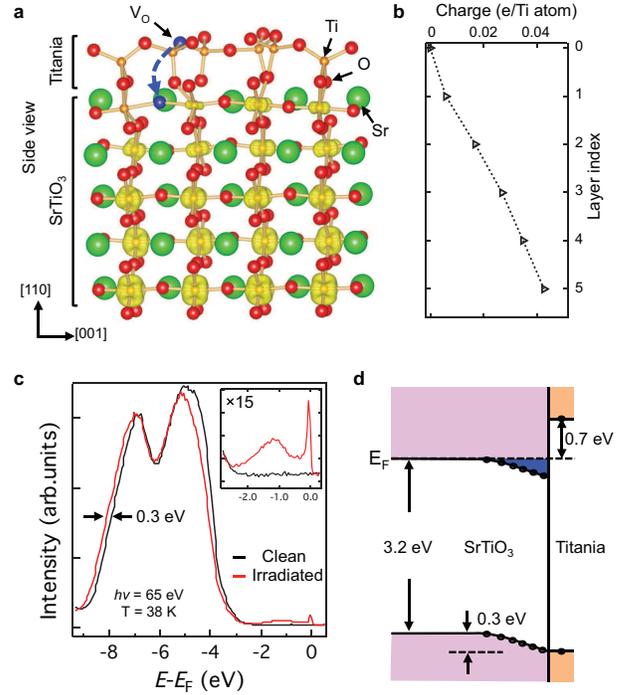}
 \caption{
{\bf SrTiO$_3$(110)-(4~$\times$~1) surface.} {\bf a} Structural model; an oxygen vacancy (V$_{\mathrm{O}}$) formed at the reconstructed surface spontaneously migrates to the (SrTiO)$^{4+}$ plane beneath the top titania layer. The excess electrons from the V$_{\mathrm{O}}$ form a 2DEG confined within a region of about 2 nm thickness. The layer-dependent charge is represented by the yellow lobes and plotted in {\bf b}. {\bf c} Angle-integrated photoemission spectroscopy of the clean surface and after  creating V$_{\mathrm{O}}$'s by synchrotron radiation.  An in-gap state and a metallic peak near the Fermi level ($E_{\mathrm{F}}$) develop, see inset. {\bf d} Schematic band structure at a surface with V$_{\mathrm{O}}$'s; the bands bend downward by 0.3 eV as deduced from the spectra in {\bf c}.  The dots denote the surface potential obtained from the shift of Ti $3s$ states in DFT+U calculations. 
}
\label{Fig1}
\end{figure}
\begin {figure*}[t]
 \includegraphics [width=6.8 in,clip] {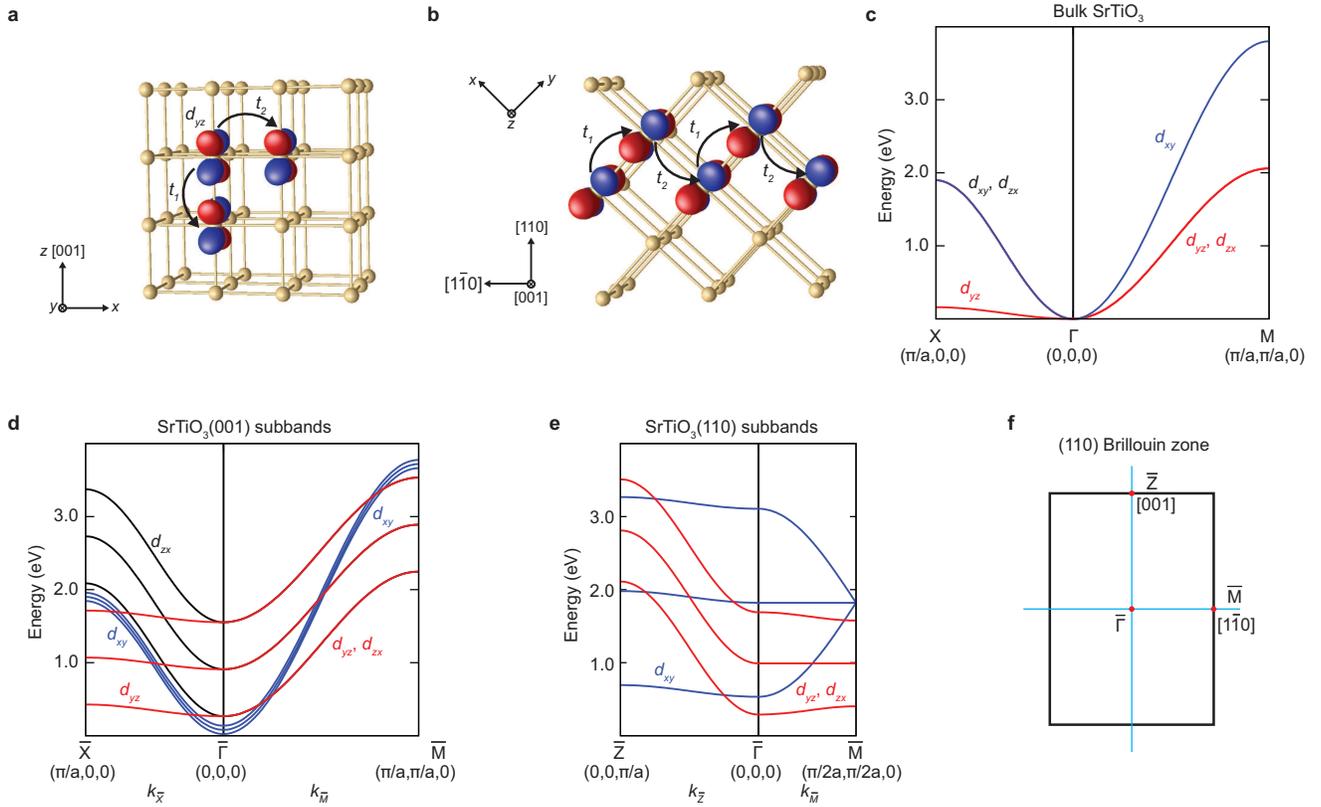}
 \caption{
{\bf Effect of Quantum confinement on the electronic structure of (001)- and (110)-oriented SrTiO$_3$.} {\bf a}-{\bf b}, Schematics of the Ti lattice in SrTiO$_3$ oriented along the [001] and [110] direction, respectively. Ti 3$d_{yz}$ orbital lobes expand in the $y$-$z$ plane. Large ($t_1$) and small ($t_2$) hopping amplitudes depend on the overlap of the nearest neighbor $d$-orbitals {\bf c}, Bulk band structure of SrTiO$_3$, consisting of a heavy $d_{yz}$ band (red) and light $d_{xy}$/$d_{zx}$ bands (blue) along $k_x$, as well as a light $d_{xy}$ band (blue) and semi-light $d_{yz}$/$d_{zx}$ bands (red) along $k_M$. {\bf d}-{\bf e}, Quantum well states (or subbands) of SrTiO$_3$ confined along [001] and [110] direction, respectively. The band dispersions of all the quantum well states confined in the (001) direction are the same as in the bulk. Confinement along (110) is different: here  the $d_{yz}$/$d_{zx}$ band becomes ``semi-heavy'' along the [1$\overline1$0] directions and the different (110)-quantum-confined states have a different mass. {\bf f}, 2D Brillouin zone of the SrTiO$_3$(110) surface.
}
\label{Fig2}
\end{figure*}

Here, we show that a 2DEG can also be induced at SrTiO$_3$(110), which is stabilized and protected by a reconstructed overlayer. This overlayer automatically forms to compensate the intrinsic polarity of the system.  A SrTiO$_3$ crystal can be viewed as a stack of alternating (SrTiO)$^{4+}$ and (O$_2$)$^{4-}$ planes along the [110] orientation,  resulting in a dipole moment that diverges with increasing crystal thickness \cite{Noguera:rpp08}.  As is often true for polar surfaces, this is prevented by one of several compensation mechanisms \cite{Noguera:rpp08}. Specifically, the  SrTiO$_3$(110) surface spontaneously forms a (4~$\times$~1) reconstruction upon various different sample treatments, including annealing in a tube furnace with flowing high-purity oxygen  \cite{Enterkin:natm10} and standard UHV  preparation procedures  \cite{Russell:prb08,Wang:prb11}.
The reconstruction consists of  a two-dimensional, tetrahedrally-coordinated titania overlayer (see Fig.~\ref{Fig1}a), which, with a nominal stoichiometry of (Ti$_{1.5}$O$_4$)$^{2-}$, quenches the overall dipole moment  \cite{Enterkin:natm10,Li:prl11}.  Because the Ti atoms in the tetrahedral titania surface layer of the reconstruction are saturated by strong, directional bonds, the  (4~$\times$~1) surface is chemically quite inert  \cite{Wang:jpcc13}.

Exposing the SrTiO$_3$(110)-(4~$\times$~1)  surface to synchrotron
radiation creates oxygen vacancies (V$_{\mathrm{O}}$'s) at the surface \cite{Wang:jpcc13}; 
they spontaneously
migrate beneath the titania overlayer, see Fig.~\ref{Fig1}a for a sketch
and Sec. S4 of the Supplementary Information for details. (This is a major 
difference to the  SrTiO$_3$(100) surface, where oxygen vacancies remain 
at the surface; hence in that case the V$_{\mathrm{O}}$'s are not protected by an overlayer and will be filled when oxygen is present at the ambient.)
The subsurface V$_{\mathrm{O}}$'s  at SrTiO$_3$(110) lead to electron-doping, and the photoemission
spectrum  in Fig.~\ref{Fig1}c shows the development of a metallic peak at E$_F$.
 Simultaneously, the O $2p$ valence band  in   Fig.~\ref{Fig1}c shifts to higher binding energy, indicating a downward band bending of $\sim$ 0.3 eV  (relative to E$_F$) in the vicinity to the surface. This is in agreement with the  density functional theory (DFT+U) \cite{vasp1,vasp2} calculated potential shown as (layer-resolved) dots in
\ref{Fig1}d.  Note that the topmost titania overlayer has a larger bandgap so that it
is not only chemically but also electrically inert.

 \begin {figure*}[t]
 \includegraphics [width=6.8 in,clip] {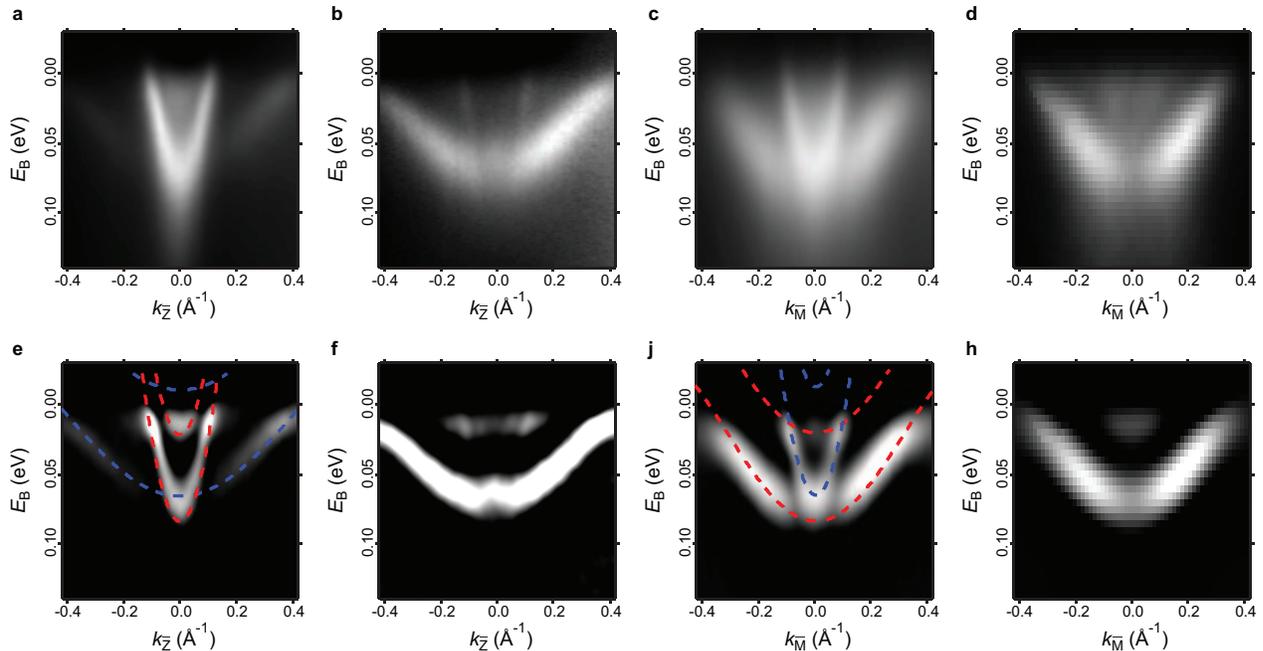}
 \caption{
{\bf ARPES of the electronic structure at SrTiO$_3$(110)-(4~$\times$~1).} {\bf a}-{\bf d}, Energy-momentum intensity maps (T$_{sample}$ = 38 K, $h\nu$ = 65 eV) along the $\overline{\Gamma}$-$\rm{ \overline{Z}}$ (or [001] ) direction and the $\overline{\Gamma}$-$\rm{ \overline{M}}$ (or [1$\overline1$0]) direction, respectively. {\bf e}-{\bf h}, Corresponding second derivatives. In each direction, the spectra were measured with linear vertical (a, d) and linear horizontal (b, c) polarized light. Tight-binding fits are overlaid for both directions. The $d_{xy}$-like bands are drawn in blue and the $d_{yz}$/ $d_{zx}$-like bands in red. The $d_{xy}$-derived bands are weakly dispersive and $d_{yz}$/$d_{zx}$-derived bands are strongly dispersive along  [001] ; the $d_{yz}$/$d_{zx}$-derived bands become weakly dispersive and $d_{xy}$-derived band becomes strongly dispersive along [1$\overline{1}$0]. The subbands become more visible in {\bf e}-{\bf h}.  
}
\label{Fig3}
\end{figure*}

The pronounced surface potential arising from the downward band banding well confines the free charge 
carriers to a thin layer so that a 2DEG develops beneath  the  titania overlayer.
 From our DFT+U calculations we conclude that the charge carriers are localized in the SrTiO$_3$ layers within about 2 nm thickness (see  Fig.~\ref{Fig1}b 
and the Supplementary Fig.~S7).

\begin {figure*}[t]
 \includegraphics [width=6.8 in,clip] {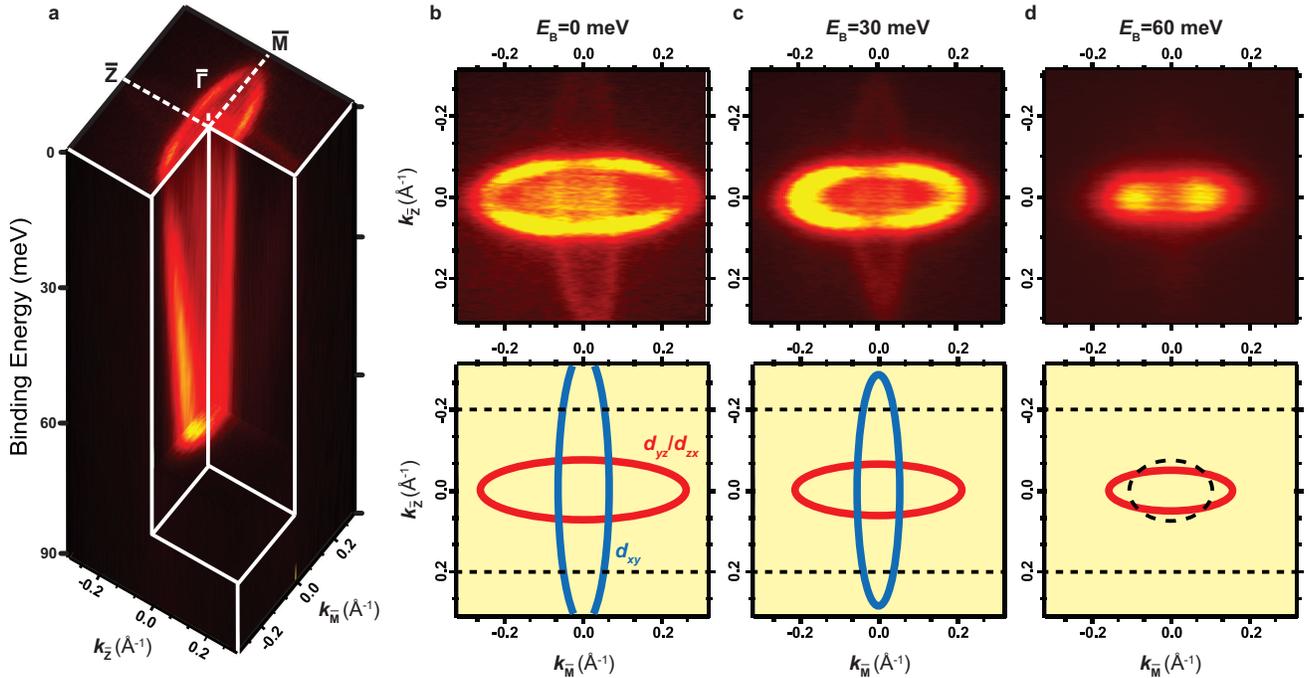}
 \caption{
{\bf Overview of the electronic structure.} {\bf a} Full photoemission mapping and {\bf b}-{\bf d} Constant energy cuts at different binding energies (E$_B$ = 0, 30 and 60 meV) and schematic constant energy surfaces (bottom panels). Data taken with LV light, which emphasizes $d_{yz}$ and $d_{zx}$ orbitals. These appear bright, while  $d_{xy}$-derived states appear faint. In the schematics the reconstructed (4~$\times$~1) Brillouin zone is indicated by dotted lines.  Note that the Fermi surface lies in the (1~$\times$~1) Brillouin zone, consistent with the 2DEG being confined at the SrTiO$_3$ layers beneath the surface reconstruction.  At higher binding energy, $E_B$ = 60 meV, only the $d_{yz}$/$d_{zx}$-derived ellipsoid is occupied. The resulting constant energy surface is strongly anisotropic compared to the bulk projected energy surface (black dashed ellipsoid).
}
\label{Fig4}
\end{figure*}

We now turn to the unusual properties of the 2DEG at  SrTiO$_3$(110)-(4~$\times$~1).
Experimentally we identify these from ARPES experiments (see Figs.~\ref{Fig3}  and \ref{Fig4}, below) but for a better understanding we start with a tight binding modeling of this 2DEG; the  details of the calculations are presented in the Sec. S8 of the Supplementary Information.
In the bulk, SrTiO$_3$ has a gap of 3.2 eV and the  three lowest conduction bands are Ti $t_{2g}$ (i.e., $d_{xy}$, $d_{yz}$ and $d_{zx}$, see Fig.~\ref{Fig2}c) orbitals that are degenerate at the $\Gamma$ point.
As the  lobes of the $d_{yz}$ orbital point into the $y$-$z$ plane (see $d_{yz}$ in Fig.~\ref{Fig2}a),  the $d_{yz}$ band has a small hopping amplitude $t_2$
in the  $\Gamma$-X direction and is hence  weakly dispersive (heavy) along   $\Gamma$-X. In contrast,  $d_{xy}$ and $d_{zx}$ have a larger overlap and 
hopping amplitude $t_{1}$ 
 in this  direction, and are thus strongly dispersive (light), as well as degenerate.   Along $\Gamma$-M, the $d_{xy}$ band is strongly dispersive (light) with hopping amplitude $t_1$, while the $d_{yz}$ and $d_{zx}$ bands are ``semi-light'' with an effective hopping amplitude ($t_1$+$t_2$)/2, see Fig.~\ref{Fig2}c.

\begin{table}[b]
\begin{ruledtabular}
\caption[Tab1]{Comparisons of experimental and theoretical effective masses of 2DEG’s. The experimental effective masses are slightly larger than those obtained from tight-binding calculations, indicating a minor mass renormalization due to electronic correlations.}
\begin{tabular}{c c c c c c c}
& \multicolumn{2}{c}{SrTiO$_3(001)$} &  \multicolumn{4}{c}{SrTiO$_3(110)$}\\
& \multicolumn{2}{c}{Along $k_{\mathrm{\overline{X}}}$} &  \multicolumn{2}{c}{Along $k_{\mathrm{\overline Z}}$} & \multicolumn{2}{c}{Along $k_{\mathrm{\overline {M}}}$}\\
\hline
\multirow{2}{*}{Exp. $m^*$($m_e$)} & \multirow{2}{*}{10$\sim$20\footnotemark[1]} & 0.7\footnotemark[1] &  \multirow{2}{*}{9.7} &  \multirow{2}{*}{0.67} &  \multirow{2}{*}{6.1} &  \multirow{2}{*}{0.74} \\
 &  & 0.5$\sim$0.6\footnotemark[2] &  &  &  &  \\
Theor. $m^*$($m_e$)& 8.2 & 0.6 & 8.2 & 0.6 & 4.7 & 0.6 \\
Orbital& $d_{yz}$ & $d_{xy}$/$d_{zx}$ & $d_{xy}$ & $d_{yz}$/$d_{zx}$ & $d_{yz}$/$d_{zx}$ & $d_{xy}$ \\ 
\end{tabular}
\label{Table I}
\end{ruledtabular}
\footnotetext[1]{Ref.\cite{Santander:nat11}}
\footnotetext[2]{Ref.\cite{Meevasana:natm11}}
\end{table}

When the  $t_{2g}$ electrons are now confined within a few nm at the SrTiO$_3$(001) surface, quantum well  states (or subbands) are formed. 
Due to the anisotropy of the $t_{2g}$ orbitals, the level spacing of the
quantized subbands  strongly depends on the orbital character \cite{Santander:nat11,Yoshimatsu:sc11,Zhong:prb13}. Notwithstanding, the [001] confinement does not change the carrier properties, i.e., the band dispersion of the subbands stays the same as in the bulk (see Fig.~\ref{Fig2}d). In striking contrast, a confinement along the [110] direction  modeled with a wedge-shaped potential well (Fig.~\ref{Fig1}d) strongly changes the properties  of the carriers, i.e., their band dispersion. 
As we show in the Supplementary Sec. S8, the effective hopping amplitude of the $d_{xy}$ orbital along the  $\overline{\Gamma}$-$\rm{ \overline{M}}$ ([1$\overline1$0]) direction becomes  $t_1$cos($\frac{\pi n}{N+1}$), where $N$ is the number of layers and $n$ the quantum number (or subband index).
That is, the effective mass of the quantum confined states now depends on the subband index  $n$.  For particular values of $n$,  such as $n=2$ in Fig.~\ref{Fig2}e, the
dispersion becomes  flat.
  This is also the case for the  $d_{yz}$/$d_{zx}$ orbitals 
where the quantum confinement leads to an
effective hopping  amplitude $\frac{2t_1t_2}{t_1+t_2}$cos($\frac{\pi n}{N+1}$) along  $\overline{\Gamma}$-$\rm{ \overline{M}}$. Here, in addition to the quantum number $n$ dependence, the prefactor also changes  from its bulk value.
As $t_2 \ll t_1$, this prefactor is $\sim 2 t_2$ so that the $d_{yz}$/$d_{zx}$ bands become ``semi-heavy'', whereas they are  ``semi-light'' ($\sim t_1/2$; Fig.~\ref{Fig2}c) in the bulk. A simple picture for this   ``semi-heavy'' behavior is that the $d_{yz}$/$d_{zx}$ electrons are quantum confined along [110] and
movement along [1$\overline1$0] requires a zig-zag path
with hopping amplitude  alternating between $t_2$ and $t_1$, see Fig.~\ref{Fig2}b. 
Hence only one $t_2$ hopping  is required for moving the electrons
 by two sites, which explains  the effective hopping amplitude  $2t_2$.
 
Figure \ref{Fig3} shows angle-resolved photoemission spectroscopy (ARPES) results from the SrTiO$_3$(110)-(4~$\times$~1) surface.  The measurements were taken along [001] and [1$\overline1$0] directions.  The intensity of the observed bands strongly depends on the light polarization (linear vertical, LV, and linear horizontal, LH). These dichroic effects are due to selection rules \cite{Aiura:ss02}, and allow disentangling the symmetry and orbital character of the various bands (see Supplementary Figs. S1-3). This enables us to ascribe the strongly and weakly dispersing bands along the [001] direction to $d_{yz}$/$d_{zx}$- and $d_{xy}$-like orbitals, respectively. Accordingly, the strongly and weakly dispersing bands along [1$\overline1$0] direction correspond to $d_{xy}$- and $d_{yz}$/$d_{zx}$-like orbitals.

The strongly (weakly) dispersing $d_{yz}$/$d_{zx}$($d_{xy}$)-like band along [001] has a bandwidth of $\sim$72 meV  ($\sim$62 meV) and a Fermi momentum of 0.11~\AA{}$^{-1}$ (0.40~\AA{}$^{-1}$). A  fit to a parabolic dispersion yields an effective mass $m^*= 0.67 m_e$ (9.7 $m_e$),  with $m_e$ the free electron mass.  These [001] effective masses are consistent with those from a 2DEG on the vacuum-fractured SrTiO$_3$(001) surface \cite{Santander:nat11,Meevasana:natm11} and the tight binding description, see Table \ref{Table I}.
 Around the $\Gamma$ point the orbital degeneracy is lifted by a  splitting  of $\sim$10 meV between the $d_{yz}$/$d_{zx}$- and $d_{xy}$-like bands.

From our tight-binding calculations we expect a quite different behavior along the  [1$\overline1$0] direction.
The $d_{yz}$/$d_{zx}$($d_{xy}$)-derived ARPES bands have bandwidths of $\sim$ 72meV (62 meV), Fermi momenta of 0.34 \AA{}$^{-1}$ (0.10 \AA{}$^{-1}$),
and  effective masses of 6.1 (0.74), see Fig.~\ref{Fig3}c,d. The former 
corresponds to a ``semi-heavy'' band  distinct from its ``semi-light''  behavior in the bulk. In particular, the ``semi-heavy'' carriers predicted by the tight-binding calculations agree well with the experimental ones, see 
Table \ref{Table I} for a summary of the effective masses.

Figure \ref{Fig3} e-h show the second derivatives of the ARPES data  along with the corresponding tight-binding subbands dispersion (Figs.~\ref{Fig3} e,j). In addition to the bands discussed above,  shallower bands becomes more visible.  These are attributed to $d_{yz}$/$d_{zx}$-like subbands, indicating that quantum well states of a 2DEG are formed at SrTiO$_3$(110). The 2D character of the observed bands is further confirmed by the fact that the bands have no dispersion along the [110] direction (surface normal) in photon energy dependent measurements (see Supplementary Fig.~S4). 
From the tight-binding calculations we found both  $d_{yz}$/$d_{zx}$- and $d_{xy}$-like carriers to be confined within $\approx$ 2 nm, in excellent agreement with the DFT+U prediction (see Supplementary Fig.~S7 and S10).

Figure \ref{Fig4} shows full photoemission mapping and constant energy cuts obtained with LV light polarization and detection along the [1$\overline1$0] direction. The resulting Fermi surface consists of two perpendicular ellipsoids and a small ellipsoid centered at the $\Gamma$ point. The bright (faint) ellipsoid is derived from $d_{yz}$/$d_{zx}$($d_{xy}$)-like band and has semi-axes of $\sim$0.11~\AA{}$^{-1}$(0.4~\AA{}$^{-1}$) and  0.34~\AA{}$^{-1}$(0.10~\AA{}$^{-1}$) along [001] and [1$\overline1$0], respectively.   From the area ($A_F$) enclosed by each Fermi surface, the corresponding 2D carrier density is $n_{2D}$ = $A_F$/2$\pi^2$. Taking into account the three bands that cross $E_F$, we find 0.39 electrons per 1~$\times$~1 unit cell of SrTiO$_3$(110) (or about 1.8~$\times$~10$^{14}$~cm$^{-2}$),  a value even larger than the sheet carrier density measured at the bare SrTiO$_3$(100) surface \cite{Santander:nat11,Meevasana:natm11}.

The Fermi surface measurements further support the conclusion that the 2DEG is not residing at, but underneath the reconstructed surface layer, as in the former case we would expect a gap opening related to the ``4$\times$'' periodicity along the [001] direction. Indeed, the Fermi surface lies in the (1~$\times$~1) but not in the reconstructed (4~$\times$~1) Brillouin zone (which is indicated by the dashed lines in Figs.~\ref{Fig4} b-d). 

Our results explain the dopant-dependent anisotropy at the (110)-oriented LaAlO$_3$/SrTiO$_3$ interface that was observed recently \cite{Annadi:natc13}.  Although both $d_{xy}$ and $d_{yz}$/$d_{zx}$-derived Fermi surfaces are strongly anisotropic, the  difference along  the [001] and  [1$\overline1$0] directions themselves is not very pronounced at higher carrier density, when both ellipsoids are occupied (see Fig.~\ref{Fig4}). However, at a higher binding energy of $E_B$ = 60 meV, only the $d_{yz}$/$d_{zx}$-derived ellipsoids appears, with a corresponding carrier density  of $\sim$~1.7~$\times$~10$^{13}$~cm$^{-2}$. Remarkably, this is the same carrier density where a pronounced anisotropic conductivity was observed in transport measurements \cite{Annadi:natc13}.   At this doping level, the big difference between the ``semi-heavy'' and light carriers comes to bear.  In the bulk, however, the anisotropy is not so pronounced (see dashed ellipsoid in Fig.~\ref{Fig4}d). 

In conclusion, we have  demonstrated that an anisotropic 2DEG with tunable properties can be created on SrTiO$_3$(110). The chemically inert and electrically insulating titania overlayer is native to this system, as it forms spontaneously to lift the intrinsic polarity of this system. It provides for a robust 2DEG that is less vulnerable against atmospheric contaminations than a 2DEG at the bare surface, and less dependent on the chemical complexity inherent to interfacial 2DEGs.  The (110) 2DEG turns out to be strikingly different from
the (001) 2DEG,  which has been the subject of previous  ARPES studies. The band dispersion is not only distinct from the one of the bulk, it even depends on the quantum number for the
(110) confinement. Hence one can engineer a completely flat band along [1$\overline1$0], which is promising for (flat band) ferromagnetism \cite{Mielke:cmp}, thermoelectricity and, possibly, superconductivity \cite{Herranz:arxiv13}.
Tuning the carrier density should provide for facile control of flat band physics as well as of the anisotopic features of the (110) 2DEG. This is possible by the amount of oxygen vacancies, applying an electric field \cite{Ueno:natm08},
and  the deposition of metal adatoms on the titania overlayer \cite{Wang:prl13}.  Our work hence paves the way for novel electronics at oxide surfaces.

{\bf \large{METHODS}}

{\bf Experiments.} The Nb-doped (0.5 wt\%) SrTiO$_3$(110) surface was prepared by cycles of Ar$^+$ sputtering (1 keV, 5 $\mu$A, 10 minutes) followed by annealing in 3~$\times$~10$^{-6}$ mbar oxygen at 900 $^{\circ}$C for 1h. The samples were heated by electron beam and the temperature was monitored with an infrared pyrometer. The surface reconstruction was checked by low energy electron diffraction (LEED) and tuned by depositing Sr metals on the surface at RT followed by annealing until a sharp (4~$\times$~1) LEED patter was observed \cite{Wang:prb11}. The ARPES measurements were performed at the ARPES 1$^2$ beamline at BESSY II, Germany. All ARPES spectra were recorded using photons with energy of 50-70 eV, linearly polarized along the horizontal/vertical direction. A Scienta R8000 analyzer with vertical detection slit geometry was used, with the energy and angular resolution of $\sim$10 meV and 0.3$^{\circ}$, respectively. Sample temperature was at $\sim$38 K.

{\bf Theory.}  DFT calculations with the inclusion of an effective on-site Coulomb repulsion U$_{eff}$ = 4.5 eV for the Ti 
$d$ states were carried out with the Vienna ab initio simulation package (VASP) \cite{vasp1,vasp2}, within the 
projector augmented-wave method and the Perdew-Burke-Ernzerhof functional \cite{pbe}.
The computational cell was modeled with a symmetric slab consisting of 45 atomic layers separated by a 
12 \AA ~thick vacuum region. One oxygen vacancy was created on both sides of the symmetrical slab (see Supplement). 
The kinetic energy cutoff for the plane-wave expansion was set to 600 eV.  We adopted a (4$\times$1) 
two-dimensional unit cell, and a (2$\times$3$\times$1) Monkhorst-Pack k-point mesh. 
During the structural optimization, atoms in the central nine layers were kept fixed to the corresponding bulk 
positions, whereas the other atoms were allowed to relax until the forces on each atom were less than 0.02 eV/\AA.

The hopping parameters for the tight binding calculations have been obtained both by (i) fitting the nearest neighbor hopping parameters to the DFT bandwidth of
bulk SrTiO$_3$ yielding $t_1=-0.455\,$ev and $t_2=-0.04\,$eV, and (ii) more thoroughly though a Wannier function projection \cite{wien2wannier,Wannier90} of a Wien2K \cite{WIEN2K} DFT calculation, using the generalized gradient approximation \cite{pbe} and 10$\times$10$\times$10 $k$-point grid. For the furthergoing tight binding calculations, up to next nearest neighbor hopping has been taken into account, see Sec. S8 and S9 of Supplementary Information for details.

{\bf \large{Acknowledgements}}
ZW, XH, and UD gratefully acknowledge support by the ERC Advanced Grant 'OxideSurfaces' and the Austrian Science Fund (FWF project F45). ZZ by the FWF through SFB VicOM F41 and KH by the ERC Starting Grant AbinitioD$\Gamma$A (grant agreement n.\ 306447). The DFT computations were performed on the Vienna Scientific Cluster (VSC-2). 


\begin{thebibliography}{10}

\expandafter\ifx\csname url\endcsname\relax
  \def\url#1{\texttt{#1}}\fi
\expandafter\ifx\csname urlprefix\endcsname\relax\def\urlprefix{URL }\fi
\providecommand{\bibinfo}[2]{#2}
\providecommand{\eprint}[2][]{\url{#2}}

\bibitem{Mannhart:sci10}
\bibinfo{author}{Mannhart, J.} \& \bibinfo{author}{Schlom, D.~G.}
\newblock \bibinfo{title}{Oxide interfaces--an opportunity for electronics}.
\newblock \emph{\bibinfo{journal}{Science}} \textbf{\bibinfo{volume}{327}},
  \bibinfo{pages}{1607--1611} (\bibinfo{year}{2010}).

\bibitem{Ohtomo:nat04}
\bibinfo{author}{Ohtomo, A.} \& \bibinfo{author}{Hwang, H.~Y.}
\newblock \bibinfo{title}{A high-mobility electron gas at the
  LaAlO$_3$/SrTiO$_3$ heterointerface}.
\newblock \emph{\bibinfo{journal}{Nature}} \textbf{\bibinfo{volume}{427}},
  \bibinfo{pages}{423--426} (\bibinfo{year}{2004}).

\bibitem{Cen:sc09}
\bibinfo{author}{Cen, C.}, \bibinfo{author}{Thiel, S.},
  \bibinfo{author}{Mannhart, J.} \& \bibinfo{author}{Levy, J.}
\newblock \bibinfo{title}{Oxide nanoelectronics on demand}.
\newblock \emph{\bibinfo{journal}{Science}} \textbf{\bibinfo{volume}{323}},
  \bibinfo{pages}{1026--1030} (\bibinfo{year}{2009}).

\bibitem{Brinkman:natm07}
\bibinfo{author}{Brinkman, A.} \emph{et~al.}
\newblock \bibinfo{title}{Magnetic effects at the interface between
  non-magnetic oxides}.
\newblock \emph{\bibinfo{journal}{Nat. Mater.}} \textbf{\bibinfo{volume}{6}},
  \bibinfo{pages}{493--496} (\bibinfo{year}{2007}).

\bibitem{Thiel:sc06}
\bibinfo{author}{Thiel, S.}, \bibinfo{author}{Hammerl, G.},
  \bibinfo{author}{Schmehl, A.}, \bibinfo{author}{Schneider, C.~W.} \&
  \bibinfo{author}{Mannhart, J.}
\newblock \bibinfo{title}{Tunable quasi-two-dimensional electron gases in oxide
  heterostructures}.
\newblock \emph{\bibinfo{journal}{Science}} \textbf{\bibinfo{volume}{313}},
  \bibinfo{pages}{1942--1945} (\bibinfo{year}{2006}).

\bibitem{Reyren:sc07}
\bibinfo{author}{Reyren, N.} \emph{et~al.}
\newblock \bibinfo{title}{Superconducting interfaces between insulating
  oxides}.
\newblock \emph{\bibinfo{journal}{Science}} \textbf{\bibinfo{volume}{317}},
  \bibinfo{pages}{1196--1199} (\bibinfo{year}{2007}).

\bibitem{Santander:nat11}
\bibinfo{author}{Santander-Syro, A.~F.} \emph{et~al.}
\newblock \bibinfo{title}{Two-dimensional electron gas with universal subbands
  at the surface of SrTiO$_3$}.
\newblock \emph{\bibinfo{journal}{Nature}} \textbf{\bibinfo{volume}{469}},
  \bibinfo{pages}{189--193} (\bibinfo{year}{2011}).

\bibitem{Meevasana:natm11}
\bibinfo{author}{Meevasana, W.} \emph{et~al.}
\newblock \bibinfo{title}{Creation and control of a two-dimensional electron
  liquid at the bare SrTiO$_3$ surface}.
\newblock \emph{\bibinfo{journal}{Nat. Mater.}} \textbf{\bibinfo{volume}{10}},
  \bibinfo{pages}{114--118} (\bibinfo{year}{2011}).

\bibitem{Plumb:arxiv13}
\bibinfo{author}{Plumb, N.~C.} \emph{et~al.}
\newblock \bibinfo{title}{Mixed dimensionality of confined conducting electrons
  tied to ferroelectric surface distortion on an oxide}.
\newblock \emph{\bibinfo{journal}{ArXiv}} \textbf{\bibinfo{volume}{1302.0708}}
  (\bibinfo{year}{2013}).

\bibitem{Bonnell:rpp08}
\bibinfo{author}{Bonnell, D.~A.} \& \bibinfo{author}{Garra, J.}
\newblock \bibinfo{title}{Scanning probe microscopy of oxide surfaces: atomic
  structure and properties}.
\newblock \emph{\bibinfo{journal}{Rep. Prog. Phys.}}
  \textbf{\bibinfo{volume}{71}}, \bibinfo{pages}{044501}
  (\bibinfo{year}{2008}).

\bibitem{Noguera:rpp08}
\bibinfo{author}{Goniakowski, J.}, \bibinfo{author}{Finocchi, F.} \&
  \bibinfo{author}{Noguera, C.}
\newblock \bibinfo{title}{Polarity of oxide surfaces and nanostructures}.
\newblock \emph{\bibinfo{journal}{Rep. Prog. Phys.}}
  \textbf{\bibinfo{volume}{71}}, \bibinfo{pages}{016501}
  (\bibinfo{year}{2008}).

\bibitem{Enterkin:natm10}
\bibinfo{author}{Enterkin, J.~A.} \emph{et~al.}
\newblock \bibinfo{title}{A homologous series of structures on the surface of
  SrTiO$_3$(110)}.
\newblock \emph{\bibinfo{journal}{Nat. Mater.}} \textbf{\bibinfo{volume}{9}},
  \bibinfo{pages}{245--248} (\bibinfo{year}{2010}).
  
 \bibitem{Russell:prb08}
\bibinfo{author}{Russell, B.} \& \bibinfo{author}{Castell, M.}
\newblock \bibinfo{title}{Reconstructions on the polar SrTiO$_3$(110) surface: Analysis using STM, LEED and AES}.
\newblock \emph{\bibinfo{journal}{Phys. Rev. B}} \textbf{\bibinfo{volume}{77}},
  \bibinfo{pages}{245414} (\bibinfo{year}{2008}).

\bibitem{Wang:prb11}
\bibinfo{author}{Wang, Z.} \emph{et~al.}
\newblock \bibinfo{title}{Evolution of the surface structures on
  SrTiO$_3$(110) tuned by Ti or Sr concentration}.
\newblock \emph{\bibinfo{journal}{Phys. Rev. B}} \textbf{\bibinfo{volume}{83}},
  \bibinfo{pages}{155453} (\bibinfo{year}{2011}).

\bibitem{Li:prl11}
\bibinfo{author}{Li, F.} \emph{et~al.}
\newblock \bibinfo{title}{Reversible transition between thermodynamically
  stable phases with low density of oxygen vacancies on the SrTiO$_3$(110)
  surface}.
\newblock \emph{\bibinfo{journal}{Phys. Rev. Lett.}}
  \textbf{\bibinfo{volume}{107}}, \bibinfo{pages}{036103}
  (\bibinfo{year}{2011}).

\bibitem{Wang:jpcc13}
\bibinfo{author}{Wang, Z.} \emph{et~al.}
\newblock \bibinfo{title}{Water adsorption at the tetrahedra titania surface
  layer of SrTiO$_3$(110)-(4$\times$1)}.
\newblock \emph{\bibinfo{journal}{J. Phys. Chem. C}}  (\bibinfo{year}{2013}).

\bibitem{vasp1}
\bibinfo{author}{Kresse, G.} \& \bibinfo{author}{Hafner, J.}
\newblock \bibinfo{title}{\textit{Ab initio} molecular dynamics for open-shell
  transition metals}.
\newblock \emph{\bibinfo{journal}{Phys. Rev. B}} \textbf{\bibinfo{volume}{48}},
  \bibinfo{pages}{13115--13118} (\bibinfo{year}{1993}).

\bibitem{vasp2}
\bibinfo{author}{Kresse, G.} \& \bibinfo{author}{Furthmüller, J.}
\newblock \bibinfo{title}{Efficiency of ab-initio total energy calculations for
  metals and semiconductors using a plane-wave basis set}.
\newblock \emph{\bibinfo{journal}{Computational Materials Science}}
  \textbf{\bibinfo{volume}{6}}, \bibinfo{pages}{15--50}
  (\bibinfo{year}{1996}).

\bibitem{Yoshimatsu:sc11}
\bibinfo{author}{Yoshimatsu, K.} \emph{et~al.}
\newblock \bibinfo{title}{Metallic quantum well states in artificial structures
  of strongly correlated oxide}.
\newblock \emph{\bibinfo{journal}{Science}} \textbf{\bibinfo{volume}{333}},
  \bibinfo{pages}{319--322} (\bibinfo{year}{2011}).

\bibitem{Zhong:prb13}
\bibinfo{author}{Zhong, Z.}, \bibinfo{author}{Zhang, Q.} \&
  \bibinfo{author}{Held, K.}
\newblock \bibinfo{title}{Quantum confinement in perovskite oxide heterostructures: Tight binding instead of a nearly free electron picture}.
\newblock \emph{\bibinfo{journal}{Phys. Rev. B}} \textbf{\bibinfo{volume}{88}}
  \bibinfo{pages}{125401} (\bibinfo{year}{2013}).

\bibitem{Aiura:ss02}
\bibinfo{author}{Aiura, Y.} \emph{et~al.}
\newblock \bibinfo{title}{Photoemission study of the metallic state of lightly
  electron-doped SrTiO$_3$}.
\newblock \emph{\bibinfo{journal}{Surf. Sci.}} \textbf{\bibinfo{volume}{515}},
  \bibinfo{pages}{61--74} (\bibinfo{year}{2002}).

\bibitem{Annadi:natc13}
\bibinfo{author}{Annadi, A.} \emph{et~al.}
\newblock \bibinfo{title}{Anisotropic two-dimensional electron gas at the
  LaAlO$_3$/SrTiO$_3$(110) interface}.
\newblock \emph{\bibinfo{journal}{Nat. Comm.}} \textbf{\bibinfo{volume}{4}},
  \bibinfo{pages}{1838} (\bibinfo{year}{2013}).

\bibitem{Mielke:cmp}
\bibinfo{author}{Mielke, A.} \& \bibinfo{author}{Tasaki, H.}
\newblock \bibinfo{title}{Ferromagnetism in the Hubbard model}.
\newblock \emph{\bibinfo{journal}{Commun. Math. Phys.}}
  \textbf{\bibinfo{volume}{158}}, \bibinfo{pages}{341--371}
  (\bibinfo{year}{1993}).

\bibitem{Herranz:arxiv13}
\bibinfo{author}{Herranz, G.}, \bibinfo{author}{Bergeal, N.},
  \bibinfo{author}{Lesueur, J.} \& \bibinfo{author}{Gazquez, J.}
\newblock \bibinfo{title}{Orientational tuning of the 2d-superconductivity in
  LaAlO$_3$/SrTiO$_3$ interfaces}.
\newblock \emph{\bibinfo{journal}{ArXiv}} \textbf{\bibinfo{volume}{1305.2411}}
  (\bibinfo{year}{2013}).

\bibitem{Ueno:natm08}
\bibinfo{author}{Ueno, K.} \emph{et~al.}
\newblock \bibinfo{title}{Electric-field-induced superconductivity in an
  insulator}.
\newblock \emph{\bibinfo{journal}{Nat. Mater.}} \textbf{\bibinfo{volume}{7}}
  (\bibinfo{year}{2008}).

\bibitem{Wang:prl13}
\bibinfo{author}{Wang, Z.} \emph{et~al.}
\newblock \bibinfo{title}{Strain-induced defect superstructure on the
  SrTiO$_3$(110) surface}.
\newblock \emph{\bibinfo{journal}{Phys. Rev. Lett.}}
  \textbf{\bibinfo{volume}{111}}, \bibinfo{pages}{056101}
  (\bibinfo{year}{2013}).

\bibitem{pbe}
\bibinfo{author}{Perdew, J.~P.}, \bibinfo{author}{Burke, K.} \&
  \bibinfo{author}{Ernzerhof, M.}
\newblock \bibinfo{title}{Generalized gradient approximation made simple}.
\newblock \emph{\bibinfo{journal}{Phys. Rev. Lett.}}
  \textbf{\bibinfo{volume}{77}}, \bibinfo{pages}{3865--3868}
  (\bibinfo{year}{1996}).

\bibitem{wien2wannier}
\bibinfo{author}{Kuneš, J.} \emph{et~al.}
\newblock \bibinfo{title}{Wien2wannier: From linearized augmented plane waves
  to maximally localized wannier functions}.
\newblock \emph{\bibinfo{journal}{Computer Physics Communications}}
  \textbf{\bibinfo{volume}{181}}, \bibinfo{pages}{1888--1895}
  (\bibinfo{year}{2010}).

\bibitem{Wannier90}
\bibinfo{author}{Mostofi, A.~A.} \emph{et~al.}
\newblock \bibinfo{title}{wannier90: A tool for obtaining maximally-localised
  wannier functions}.
\newblock \emph{\bibinfo{journal}{Computer Physics Communications}}
  \textbf{\bibinfo{volume}{178}}, \bibinfo{pages}{685--699}
  (\bibinfo{year}{2008}).

\bibitem{WIEN2K}
\bibinfo{author}{Blaha, P.}, \bibinfo{author}{Schwark, K.}, \bibinfo{author}{Madsen, G.~K.~H.}, \bibinfo{author}{Kvasnicka, D.} \&
  \bibinfo{author}{Luitz, J.}
 \emph{\bibinfo{title}{WIEN2k, An Augmented Plane Wave + Local Orbitals Program
  for Calculating Crystal Properties, edited by K. Schwarz, Technische Universit\"{a}t Wien (2001)}}.

\end{thebibliography}
\end{document}